\documentclass[12pt,reqno]{amsart}
\usepackage{amsmath,amssymb,amsthm,amscd,wasysym,mathrsfs,amsxtra,mathtools}
\input xy
\xyoption{all}

\usepackage[margin=2.3cm]{geometry}
\usepackage{enumitem}
\setlist[enumerate,1]{font=\normalfont, label=(\roman*)}

\usepackage{hyperref}

\usepackage{tikz}
\usetikzlibrary{backgrounds}

\theoremstyle{definition}

\newcommand{\IC}{\mathbb{C}}
\newcommand{\IR}{\mathbb{R}}
\newcommand{\IZ}{\mathbb{Z}}

\newcommand{\bb}{{\bf b}}
\newcommand{\bd}{{\bf d}}
\newcommand{\bt}{{\bf t}}

\title[Nonperturbative refined topological string]{Nonperturbative refined topological string}  
\author[Wu-yen Chuang]{Wu-yen Chuang}
\address{Department of Mathematics and TIMS, National Taiwan University, Taipei, Taiwan}
\email{wychuang@gmail.com}


\begin{document}

\begin{abstract} 
A formula for the full nonperturbative topological string free energy was recently proposed by Hattab and Palti \cite{HP24a}. In this work, we extend their result to the refined topological string theory. We demonstrate that the proposed formula for the full nonperturbative refined topological string free energy correctly reproduces the trans-series structure of the refined topological string and captures the Stokes automorphisms associated with its resurgent properties.
\end{abstract}

\maketitle

\section{Introduction}
The free energy of topological string theory has a perturbative expansion given by, 
\begin{equation}\label{eq:free}
	\mathcal{F}_{\text{pert}}(\lambda) = \sum_{g=0}^{\infty} \mathcal{F}_g\ {\lambda}^{2g-2} 
\end{equation} where $g$ is the genus and $\lambda$ represents the coupling constant of the string theory. At each genus it is expected that $\mathcal{F}_g$ goes as $\mathcal{F}_g \sim (2g-3)!$ due to diverging number of the contributing diagrams at higher orders. Consequently the series (\ref{eq:free}) would at best be an asymptotic series of zero convergence radius. By analyzing the breakdown of the series, it becomes evident that the free energy must be corrected into a more complete form,
\begin{equation}
	\mathcal{F}_{\text{full}}(\lambda) = \sum_{g=0}^{\infty} \mathcal{F}_g\ {\lambda}^{2g-2} + \mathcal{O}(e^{-\frac{1}{\lambda}}) \ .
\end{equation} In other words, the full partition function of the string theory is expected to be an analytic function of various parameters. Instead of limiting analysis to asymptotic expansions at specific points in parameter space, it is desirable to determine the full function.

One approach to obtaining the nonperturbative free energy is to apply 
the theory of resurgence \cite{Ecal81}. Given a divergent series 
$f(x)=\sum_{n=0}^{\infty}a_n x^n$ with coefficients scaling as 
$a_n \sim n!/A^n$ for some $A$, the Borel transform $\mathcal{B}[f]$ 
is defined as
\begin{equation}
	\mathcal{B}[f](s) = \sum_{n=0}^{\infty}\frac{a_n}{n!} s^n \ ,
\end{equation} which is a series with a singularity at $s=A$ and convergent for $|s|<|A|$. Assuming that $\mathcal{B}[f](s)$ can be continued analytically to the whole positive real line and $A \notin \IR_+$, one can verify that the Borel resummation $\mathcal{S} f(x)$ of $f(x)$, defined by
\begin{equation}
	\mathcal{S} f(x) = \int_{0}^{\infty} \mathcal{B}[f](sx) e^{-s} ds, 
\end{equation} retains the identical asymptotic expansion as the original function $f(x)$. If $A \in \IR_+$, the integration contour must be
deformed to avoid $A$, leading to two distinct Borel resummations $\mathcal{S}_{\pm} f(x)$. Resurgence theory, combined with holomorphic anomaly equations, has significantly advanced the construction of trans-series solutions for topological string theory \cite{CESV16, GKKM24} and its refined version \cite{AMP24}. See the references therein for a more comprehensive list of related works. 

Recent works \cite{HP24a, HP24b} have provided an alternative approach, deriving the full nonperturbative free energy from an integrating-out calculation of M2-branes. Their final expression differs from the original Gopakumar-Vafa computation \cite{GV} due to subtleties arising from nonperturbative pole structures.

The key point of \cite{HP24a}\cite{HP24b} is that performing the integrating-out calculation carefully yields the following expression for the free energy of the topological string or Gromov-Witten theory on a Calabi-Yau threefold $X$,

\begin{equation}\label{eq:contour1}
	\mathcal{F}_{\text{full}}=\sum_{\beta, g \geq 0, n \in \IZ} \alpha_{g}^{\beta} \int_{\epsilon}^{\infty} \frac{ds}{s} \frac{e^{-s |z_{\beta,n}|^2}}{(2 \sin (\bar{z}_{\beta,n} s \lambda /2))^{2-2g}} \ ,
\end{equation} where $\alpha_{g}^{\beta}$ is Gopakumar-Vafa invariant, {\it i.e.} the degeneracy number of the BPS states with central charge $z_{\beta,n}=\beta \cdot \bt - 2 \pi i n$ and genus $g$, $\beta \in H_2(X, \IZ)$ is an effective curve class in $X$, and $\bt$ is the complexified K\"ahler form.

Moreover, the integral representation (\ref{eq:contour1}) naturally captures the Stokes jumps predicted by resurgence theory. Specifically, the Stokes automorphisms governing Borel resummation discontinuities correspond to shifts in nonperturbative corrections when the integration contour crosses singularities. In this framework, the Stokes constants can be directly identified with Gopakumar-Vafa invariants.

These insights lead to the main goal of this paper: extending the contour integral formulation of the full nonperturbative free energy (\ref{eq:contour1}) to refined topological string theory.
After revisiting the computations by Gopakumar-Vafa \cite{GV}, Dedushenko-Witten \cite{DW16} and \cite{HP24b}\cite{HP24b} in the $\Omega$-background
and comparing with the various inputs from \cite{AMP24}\cite{CDP14}, we propose that the full nonperturbative {\it refined} topological string free energy $\mathcal{F}_{\text{ref,full}}$ takes the form,
\begin{equation}\label{eq:refinedcontour1}
	\mathcal{F}_{\text{ref,full}}=\sum_{\stackrel{\beta, j_L, j_R}{n \in \IZ}} N^{\beta}_{j_L,j_R} \int_{\epsilon}^{\infty} \frac{ds}{s} \frac{e^{-s |z_{\beta,n}|^2}\ 
		\chi_{j_L}(e^{i \bar{z}_{\beta,n} s\lambda(\bb+\bb^{-1})/2})\ \chi_{j_R}(e^{i \bar{z}_{\beta,n} s\lambda(\bb-\bb^{-1})/2})}{4 \sin (\bar{z}_{\beta,n} s \bb \lambda /2)\sin (\bar{z}_{\beta,n} s \lambda / 2\bb)} \ ,
\end{equation} where $\bb$ is the refinement parameter, $\chi_j(y)=(y^{2j+1}-y^{-2j-1})/(y-y^{-1})$ is the character of the spin $j$ representation of $SU(2)$, 
and $N^{\beta}_{j_L,j_R}$ counts BPS states with effective class $\beta \in H_2(X,\IZ)$ and spin quantum number $(j_L, j_R) \in \IZ/2 \times \IZ/2$ under the little group $SU(2)_L \times SU(2)_R$  in five dimensions. 

Further analysis of (\ref{eq:refinedcontour1}) confirms that this expression correctly reproduces the perturbative refined topological string free energy, as well as the trans-series structure and Stokes automorphisms described in \cite{AMP24}.

The paper is organized as follows. In Section 2, following \cite{AMP24}, we provide a concise review of refined topological string theory, covering trans-series solutions and the Stokes automorphisms associated with Borel singularities. In Section 3, we derive an integral representation for the full nonperturbative refined topological string free energy by analyzing M2 BPS states in the $\Omega$-background. In Section 4, we demonstrate that this new formula reproduces the trans-series structure of the refined topological string and captures the Stokes automorphisms arising from resurgence. Finally, in Section 5, we summarize our findings and outline potential directions for future research.

\section{The refined topological string}\label{refinedtop}

In this section we collect some basic properties of the refined topological string theory and its trans-series and the Stokes automorphisms from \cite{AMP24}.

\subsection{The perturbative refined topological string}

The refined topological string has two complex parameters $\epsilon_1$ 
and $\epsilon_2$. They are related to the string coupling $\lambda$ by 
\begin{equation}
	\epsilon_1= \lambda \bb, \quad \epsilon_2= -\lambda \bb^{-1}.
\end{equation} Setting $\bb=1$ recovers the topological string
theory. The perturbative part of the refined topological string free energy 
is given by 
\begin{equation}
	\mathcal{F}(\bt, \epsilon_1, \epsilon_2) = \sum_{i+j \geq 0}
	(\epsilon_1+\epsilon_2)^{2i}(-\epsilon_1 \epsilon_2)^{j-1}
	\mathcal{F}^{(i,j)}({\bf t}), 
\end{equation} where $\bt$ are the flat coordinates on the moduli space of $X$. In A-model, they are complexified K\"ahler class. 
We can also expand the pertubative free energy in the string coupling $\lambda$
such that the coefficients depend on the moduli $\bt$ and the deformation $\bb$,
\begin{equation}
	\mathcal{F}(\bt, \lambda, \bb) = \sum_{g \geq 0} \lambda^{2g-2} \mathcal{F}_g(\bt, \bb), 
\end{equation} where 
\begin{equation}
\mathcal{F}_g(\bt, \bb) = \sum_{k=0}^{g} 
\mathcal{F}^{(k, g-k)}({\bf t})
(\bb-\bb^{-1})^{2k}.
\end{equation} The refined free energy $\mathcal{F}_g(\bt, \bb)$
can be computed on certain local CY geometries using instanton calculus \cite{CDP14}
or refined topological vertex techniques on A-model side \cite{IKV09}. The B-model approach to the refined case is based on the refined holomorphic anomaly equations \cite{HK12}. 

\medskip
Generalizing the Gopakumar-Vafa integrality structure to the refined case, we have the following expression for the 
perturbative refined free energy in term of refined Gopakumar-Vafa or BPS invariants, 

\begin{equation}
\mathcal{F}(\bt, \lambda, \bb) = 
\sum_{k, {\bf d}} \sum_{j_L, j_R}	
\frac{N^{\bd}_{j_L, j_R}}{k} \frac{ \chi_{j_L}(e^{i k (\epsilon_1-\epsilon_2)/2})\ \chi_{j_R}(e^{i k (\epsilon_1+\epsilon_2)/2})}{4 \sin ( k \lambda \bb /2) \sin ( k \lambda / 2\bb)} 
 e^{-k \bd \cdot \bt}, 
\end{equation} where $N^{\bd}_{j_L, j_R}$ are integers, counting BPS states with charge $\bd$ transforming with spin quantum number 
$(j_L, j_R) \in \IZ^+/2 \times \IZ^+/2$ under the little group $SU(2)_L \times SU(2)_R$ in five dimensions. When $\bb=1$ the conventional Gopakumar-Vafa integrality structure of the topological string free energy is recovered. In particular we have the relation between the genus zero GV invariants $\alpha^{\bd}_0$ and refined BPS invariants 
$N^{\bd}_{j_L, j_R}$, 
\begin{equation}
\alpha^{\bf d}_0 = \sum_{j_L, j_R} (2 j_L+1)(2 j_R+1)N^{{\bf d}}_{j_L, j_R},
\end{equation} which comes from the coefficient of $\lambda^{-2}$.

Notice that we use the notation $N^{\bd}_{j_L, j_R}$ or
$N^{\beta}_{j_L, j_R}$ interchangeably, where the effective class $\beta \in H_2(X,\IZ)$ and the tuple $\bd$ refers to expanding $\beta$ in terms of a basis. 

The refined Donaldson-Thomas invariants $\Omega({\bf d},y)$ for the class $\beta$ or the tuple $\bd$ are the characters for the diagonal $SU(2) \subset SU(2)_L \times SU(2)_R$,
\begin{equation}
	\Omega(\bd,y)=\sum_{j_L, j_R} \chi_{j_L}(y)\chi_{j_R}(y) 
	N^{\bd}_{j_L, j_R}.
\end{equation}

The refined DT invariants are in fact defined for any general class $\gamma \in K_0(X)$ and can be decomposed in terms of $SU(2)$ characters,
\begin{equation}
	\Omega(\gamma, y) = \sum_j \chi_{j}(y) \Omega_{[j]}(\gamma), 
\end{equation} where $\Omega_{[j]}(\gamma)$ counts the BPS states of charge $\gamma$ and spin quantum number $j$. 

\medskip

As already alluded above, on local toric Calabi-Yau threefolds, the refined free energy $\mathcal{F}_g(\bt, \bb)$ can be computed via refined topological vertex formalism on A-model side \cite{IKV09}, and through the refined holomorphic anomaly equations 
on B-model side \cite{HK12}. The refined topological string could also be defined on compact Calabi-Yau threefolds \cite{HKK21}, but the free energies will acquire dependence on complex structure moduli. 
More recently 
a novel set of refined holomorphic anomaly equations for compact Calabi-Yau threefolds has been derived \cite{HKKW25} to compute 
the refined BPS invariants. 

The mathematical definition for Gopakumar-Vafa or BPS invariants is based
on the perverse sheaves of vanishing cycles on moduli spaces of 
one dimensional sheaves (D2-D0 branes) on the Calabi-Yau threefolds \cite{MT18}.  It is natural to expect that the perverse cohomology of the perverse sheaves of vanishing cycles may lead to a refined version of GV invariants. However, a rigorous mathematical proof of this connection is still lacking.

Another significant mathematical proposal for the refined 
topological string theory on local toric Calabi-Yau threefolds was made
by Nekrasov and Okounkov. 
In \cite{NO16} they introduced an index computation on an eleven dimensional M-theory background $\mathcal{M}_{11}=\mathcal{W}_{10} \times \IR_{\text{time}} $ fibered over a six dimensional manifold $X$. For our purpose $X$ is taken to be the local toric Calabi-Yau threefold under consideration. Here $\mathcal{W}_{10}$ is a splitting 
rank two holomorphic vector bundle over $X$, equipped with a $k$ centered Taub-Nut metric and a $U(1) \subset U(2)$ isometry. 
Upon reducing M-theory along the $U(1)$ circle, one obtains IIA theory on
$\mathcal{M}_{10}=\det{\mathcal{W}} \times \IR^{1,1}$ over $X$, where 
$\det{\mathcal{W}}$ denotes the determinant line bundle.
This geometry models $k$ D6 branes wrapping on $X$. In the case $k=1$ and $X$ local toric, the results obtained from the refined topological vertex formalism were reproduced in \cite{NO16}. 

\subsection{Large genus behaviors of the refined topological string and trans-series solutions}

Define the coefficients $\sigma_g^{j_L, j_R}(\bb)$ by the following expansion,
\begin{equation}
	\frac{ \chi_{j_L}(e^{i x \bb_+/2})\ \chi_{j_R}(e^{i x \bb_-/2})}{4 \sin ( x \bb /2) \sin ( x / 2 \bb)} = 
	\sum_{g \geq 0} \sigma_g^{j_L, j_R}(\bb) 
	x^{2g-2}, \quad \bb_\pm = \bb \pm \bb^{-1}.
\end{equation} 

Then the genus $g$ perturbative refined free energy $\mathcal{F}_g(\bt, \bb)$ is given by
\begin{equation}
	\mathcal{F}_g(\bt, \bb) = \sum_{\bd, j_L, j_R} N^\bd_{j_L, j_R} 
	\sigma_g^{j_L, j_R}(\bb) \text{Li}_{3-2g}(e^{-\bd \cdot \bt}).
\end{equation}

Define the following quantities,
\begin{align}
	& A_{\bd,n}= 2 \pi \bd \cdot \bt + 4 \pi^2 i n, \\ 
	& y_\bb = - e^{\pi i \bb^2}, \  \tilde{y}_\bb = - e^{-\pi i / \bb^2}.
\end{align}

By analyzing the asymptotic expansion of $\sigma_g^{j_L, j_R}(\bb)$ for 
$g \gg 1$, it was concluded in \cite{AMP24} that there are Borel 
singularities at 
\begin{equation}
	\frac{l}{\bb} A_{\bd,n}, \quad l \bb A_{\bd,n}, \quad l \in \IZ.
\end{equation} 

The trans-series corresponding to the singularities are given by
\begin{align}
	& \mathcal{F}^l_{A_{\bd,n},\bb,j_L, j_R} = \frac{(-1)^{l-1}}{l} \frac{ \chi_{j_L}(\tilde{y}_\bb^l)\ \chi_{j_R}(\tilde{y}_\bb^l)}{2 \sin ( \pi l / \bb^2)} e^{-\frac{l A_{\bd,n}}{\bb \lambda}}, \label{eq:tseries} \\
	& \mathcal{F}^l_{A_{\bd,n},\bb^{-1},j_L, j_R} = \frac{(-1)^{l-1}}{l} 
	\frac{ \chi_{j_L}(y_\bb^l)\ \chi_{j_R}(y_\bb^l)}{2 \sin ( \pi l \bb^2)} e^{-\frac{l \bb A_{\bd,n}}{\lambda}} .
	\label{eq:tseries'}
\end{align}

\subsection{Interlude: quantum dilogarithms}
In this subsection we take a necessary digression to introduce 
the quantum dilogarithm functions \cite[Appendix A]{AMP24}. 

The quantum dilogarithm function $E_y(x)$ is defined for $x,y \in \IC$ as
\begin{equation}
	E_y(x)= \exp \Big[\sum_{k=1}^{\infty} \frac{(xy)^k}{k(1-y^{2k})}\Big]=
	\prod_{n=1}^{\infty} (1-xy^{2n+1})^{-1},
\end{equation} where the product converges for $|y|<1$.

Yet another version of quantum dilogarithm was introduced by Faddeev \cite{Fad95}, as an integral representation,
\begin{equation}\label{eq:Fadqdilog}
	\Phi_\bb(z) = \exp \big(\int_{\IR+i\epsilon} 
	\frac{e^{-2izs}}{4 \sinh (s\bb) \sinh(s/\bb)} \frac{ds}{s} \big)
\end{equation} where the integration is over the real line, avoiding the 
pole at $v=0$ by deforming the contour into the upper half plane.

After extracting out the residues in (\ref{eq:Fadqdilog}) and taking $\log$, we obtain
\begin{equation}
	\log \Phi_\bb(z)=\sum_{l=1}^{\infty} \frac{(-1)^l}{2 i l} 
	\Big(\frac{2 \pi z \bb}{\sin (l\pi\bb^2)} +
	\frac{2 \pi z /\bb}{\sin (l\pi/\bb^2)}\Big) . 
\end{equation}

In \cite{FKV01}, two different versions of quantum dilogarithm functions are related by,
\begin{equation}
	\Phi_\bb(z) = \frac{E_{\tilde{y}_\bb}(e^{2\pi z /\bb})}
	{E_{y_\bb}(e^{2\pi z \bb})} , \quad \text{for }\Im(\bb^2)>0.
\end{equation} 

We introduce another generalization of $E_y(z)$ associated 
with the spin $j$ representation,
\begin{equation}
	E_y^{[j]}(x) = \exp \Big[\sum_{k=1}^{\infty} 
	\frac{(xy)^k \chi_j(y^k)}{k(1-y^{2k})}\Big],
\end{equation} where $\chi_j$ is $SU(2)$ character of spin $j$.
One can verify that $E_y^{[j]}(x)$ is a product of $E_y$ with 
different arguments,
\begin{equation}
	E_y^{[j]}(x) = \prod_{m=-j}^{j} E_y(xy^{2m}).
\end{equation} We define the corresponding generalization for 
$\Phi_\bb(z)$ as follows,
\begin{equation}
	\Phi_\bb^{[j]}(z) = \frac{E_{\tilde{y}_\bb}^{[j]}(e^{2\pi z /\bb})}
	{E_{y_\bb}^{[j]}(e^{2\pi z \bb})}.
\end{equation}

It follows that $\Phi_\bb^{[j]}(z)$ has the following expressions,

\begin{align}
	\Phi_\bb^{[j]}(z) &= \exp\Big( \sum_{l=1}^{\infty} \frac{(-1)^l}{2 i l} \Big( \frac{\chi_j(-e^{\pi i l \bb^2}) e^{2 \pi l z \bb}}{\sin(l \pi \bb^2)} 
	+\frac{\chi_j(-e^{-\pi i l / \bb^2}) e^{2 \pi l z/ \bb}}{\sin(l \pi /\bb^2)} \Big)
	\Big) \nonumber \\
	&= \exp\Big( \int_{\IR + i\epsilon} \frac{\chi_j(e^{(\bb-\bb^{-1})s}) e^{-2izs}}
	{4 \sinh(s\bb) \sinh(s/\bb)} \frac{ds}{s} \Big).
\end{align}

\subsection{Trans-series solutions}

Now we are ready to sum up the trans-series (\ref{eq:tseries})
(\ref{eq:tseries'}), 
corresponding to the Borel singularities $\frac{l}{\bb} A_{\bd,n}, \ l \bb A_{\bd,n}$ respectively.

The sum over $l$ is given by 
\begin{equation}\label{eq:sum_tseries}
	\sum_{l \geq 1} \big( \mathcal{F}^l_{A_{\bd,n},\bb,j_L, j_R} + \mathcal{F}^l_{A_{\bd,n},\bb^{-1},j_L, j_R} \big)
	= \sum_{l \geq 1} \frac{(-1)^{l-1}}{l}\Big[\frac{ \chi_{j_L}(\tilde{y}_\bb^l)\ \chi_{j_R}(\tilde{y}_\bb^l)}{2 \sin ( \pi l / \bb^2)} e^{-\frac{l A_{\bd,n}}{\bb \lambda}} + 
	\frac{ \chi_{j_L}(y_\bb^l)\ \chi_{j_R}(y_\bb^l)}{2 \sin ( \pi l \bb^2)} e^{-\frac{l \bb A_{\bd,n}}{\lambda}} \Big] .
\end{equation} 

Define $\Phi_\bb^{[j_L,j_R]}(z)$ in terms of 
$\Phi_\bb^{[j]}(z)$,
 
\begin{equation}
	\log \Phi_\bb^{[j_L,j_R]}(z) := \sum_{j=|j_L-j_R|}^{j_L+j_R}
	\log \Phi_\bb^{[j]}(z), 
\end{equation} where the sum over $j$ runs over all half-integers such that $j-j_L-j_R$ is integer.

Therefore, (\ref{eq:sum_tseries}) can be simplified to,
\begin{equation}
	\sum_{l \geq 1} \big( \mathcal{F}^l_{A_{\bd,n},\bb,j_L, j_R} + \mathcal{F}^l_{A_{\bd,n},\bb^{-1},j_L, j_R} \big)
	= -i \log \Phi_\bb^{[j_L,j_R]}\Big(-\frac{A_{\bd,n}}{2 \pi \lambda}\Big) .
\end{equation}

Summing over $\bd, n, j_L, j_R$ along with the degeneracy number $N^{{\bf d}}_{j_L, j_R}$, we obtain 
the full trans-series near large radius,
\begin{equation}\label{eq:tseries2}
	-i \sum_{\bd,n\in \IZ, j_L,j_R} N^{{\bf d}}_{j_L, j_R} \log \Phi_\bb^{[j_L,j_R]}
	\Big( -\frac{A_{\bd,n}}{2 \pi \lambda}\Big) = -i \sum_{\bd,n\in \IZ, j} 
	\Omega_{[j]}(\bd) \log \Phi_\bb^{[j]} \Big( -\frac{A_{\bd,n}}{2 \pi \lambda}\Big),
\end{equation} where
	
\begin{equation}
	\Omega_{[j]}(\bd) = \sum_{\stackrel{j_L,j_R}{|j_L-j_R|\leq j \leq j_L+j_R}} N^{{\bf d}}_{j_L, j_R}.
\end{equation}

\subsection{Stokes automorphism in refined topological strings}
It was found in \cite{AMP24} that for the refined free energy at large radius, the
leading Borel singularities are at $l \bb^{\pm 1} A_{\bd,n}, l \in \IZ\setminus 0$, 
corresponding to D2-D0 branes. For general $\bb$ we have two rays of singularities.  
We consider the Stokes automorphisms associated with the discontinuity when crossing 
two rays for a given $(\bd,n)$. 
We have 
\begin{equation}
	\mathfrak{S}(\mathcal{Z}) = \prod_{j_L,j_R} \Big[ \Phi_\bb^{[j_L,j_R]}
	\Big(-\frac{A_{\bd,n}}{2 \pi \lambda}\Big) \Big]^{-N^{\bd}_{j_{L},j_R}} \mathcal{Z}
	= \prod_{j} \Big[ \Phi_\bb^{[j]} \Big(-\frac{A_{\bd,n}}{2 \pi \lambda }\Big)\Big]
	^{-\Omega_{[j]}(\bd)} \mathcal{Z},
\end{equation} where $\mathcal{Z}$ is the partition function.
In Section 4, we will reproduce these Stokes jumps from the
perspective of an integral formula for the 
full nonperturbative refined topological string free energy. 

\section{Nonperturbative refined topological string}

In \cite{HP24a}\cite{HP24b} the authors examined the integrating-out calculation of the M2-branes in \cite{GV}\cite{DW16} and obtained the expression for the full free energy
\begin{equation}\label{eq:full}
	\mathcal{F}_{\text{full}}=\sum_{\beta, g \geq 0, n \in \IZ} \alpha_{g}^{\beta} \int_{\epsilon}^{\infty} \frac{ds}{s} \frac{e^{-s |z_{\beta,n}|^2}}{(2 \sin (\bar{z}_{\beta,n} s \lambda /2))^{2-2g}} \ ,
\end{equation} where $\alpha_{g}^{\beta}$ is the degeneracy number of the BPS states with central charge $z_{\beta,n}=\beta \cdot \bt - 2 \pi i n$ and genus $g$, $\beta \in H_2(X, \IZ)$ is an effective curve class in $X$, and $\bt$ is the complexified K\"ahler form. 
The factor $(2 \sin (\bar{z}_{\beta,n} s \lambda /2))^2$ in the denominator comes from integrating out M2 states in the presence of anti self-dual graviphoton field strength, which is identified with topological string coupling constant $\lambda$. 

Let $W_{\mu\nu}$ is the graviphoton field strength in 4 dimensions. And $W_{\mu\nu}^+$ and $W_{\mu\nu}^-$ are the self-dual and anti self-dual parts of $W_{\mu\nu}$, 
\begin{equation}
	W_{\mu\nu}^{\pm} = \frac{1}{2} \big(W_{\mu\nu} \pm \frac{1}{2} \epsilon_{\mu\nu\rho\lambda}W^{\rho\lambda}\big). 
\end{equation} In \cite{HP24b} the anti self-dual condition on the graviphoton field strength is imposed by setting 
$W_{12}^{-}=-W_{34}^{-}$.

Now we review some key steps in the derivation of (\ref{eq:full}) in \cite[Section 4]{HP24b}.
One useful and analogous computation is integrating out a 
charged scalar in four Euclidean dimensions. 
Consider the four dimensional scalar QED in Euclidean 
signature,
\begin{equation}
	S^E_{QED} = \int d^4x \Big[\frac{1}{4} F_{\mu \nu}F^{\mu \nu} + \phi^*(-D^2+m^2) \phi \Big],
\end{equation} where $D_{\mu} = \partial_\mu - i e A_\mu$. 
Define the effective action $S_{\text{eff}} = \frac{1}{4}F_{\mu \nu}F^{\mu \nu} + \Delta S[A] $ by the 
path integral,
\begin{equation}
	\int \mathcal{D}A\ \mathcal{D}\phi^* \mathcal{D}\phi\ e^ {-S^E_{QED}[A,\phi]} = \int \mathcal{D}A\ e^{-S_{\text{eff}}}.
\end{equation}

Using the Schwinger's representation for an operator $\mathcal{N}$,
\begin{equation}
	\frac{1}{\mathcal{N}}=\int_{0}^{\infty} ds \ e^{-s{N}},
\end{equation} we obtain
\begin{equation}
	\Delta S[A] = - \int_{\epsilon}^{\infty} \frac{ds}{s}
	e^{-s m^2} \text{Tr}(e^{-sH}), \quad H= -D^2,
\end{equation} where the cutoff $\epsilon$ is introduced 
to absorb the integration constant and infinite normalization
constant. The result can be applied to the integrating out 
computation for the fields associated to the wrapped M2 branes.
The scalar fields in consideration have mass $|z_{\beta,n}|$
and are coupled to the graviphoton.
Since $W_{12}^{-}=-W_{34}^{-}$, we have
\begin{equation}
	\text{Tr}(e^{sD^2}) = 
	\text{Tr}(e^{s(D_1^2+D_2^2)})\text{Tr}(e^{s(D_3^2+D_4^2)}).
\end{equation} 
And each trace factor contributes $2 \sin (\bar{z}_{\beta,n} s \lambda /2)$ to the denominator, where $\lambda$ is proportional to $|W_{12}^-|$. Another $2g$ factors come from 
taking trace over the Hilbert space $\mathcal{H}_g$ of $g$ 
hypermultiplets,
\begin{equation}
	\text{Tr}_{\mathcal{H}_g} (-1)^F e^{-s \widetilde{H}} ,
\end{equation} where $\widetilde{H} \sim 2i \bar{z}_{\beta,n} \lambda J$, $J$ is the helicity operator, and the fermion number $F=2J$. Here we use the notation $\sim$ since the identification of $\tilde{H}$ involves certain parameter redefinitions. 

Combining all the contributions and taking into account the degeneracy $\alpha_g^{\beta}$, we arrive at the formula 
(\ref{eq:full}).
 
\medskip
For the refined topological string, one needs to consider the $\Omega$ background and turn on the graviphoton field strengh in the following way,
\begin{equation}
	W_{12}^{-} \sim \bb \lambda, \quad 
	W_{34}^{-} \sim -\lambda/\bb. 
\end{equation} 

As in the conventional anti self-dual case \cite[Section 4]{HP24b}, we integrate out M2 branes wrapping on two cycles in the Calabi-Yau threefold. In this case, the trace factor $\text{Tr}(e^{sD^2})$ is modified to 
\begin{equation}
4 \sin (\bar{z}_{\beta,n} s \bb \lambda /2)\sin (\bar{z}_{\beta,n} s \lambda / 2\bb).
\end{equation}
For hypermultiplets, we organize the states according to 
the representation $(j_L,j_R)$ under the little group
$SU(2)_L \times SU(2)_R$.
Due to the grviphoton field strength, we take the trace
over the Hilbert space $\mathcal{H}_{j_L, j_R}$,
\begin{equation}
	\text{Tr}_{\mathcal{H}_{j_L, j_R}} 
	(-1)^F e^{-s \widetilde{\widetilde{H}}}, 
\end{equation} where 
$\widetilde{\widetilde{H}} \sim i \bar{z}_{\beta,n} \big(
(\bb+\bb^{-1})\lambda J_L + (\bb - \bb^{-1})\lambda J_R\big)
$, $J_L$ and $J_R$ are the helicity operators, and the fermion number $F=2(J_L+J_R)$.
The trace over a single $(j_L,j_R)$ representation gives
\begin{equation}
		\text{Tr}_{\mathcal{H}_{j_L, j_R}} 
	(-1)^F e^{-s \widetilde{\widetilde{H}}}=
	(-1)^{j_L+j_R}\chi_{j_L}(e^{i \bar{z}_{\beta,n} s\lambda(\bb+\bb^{-1})/2})\ \chi_{j_R}(e^{i \bar{z}_{\beta,n} s\lambda(\bb-\bb^{-1})/2}), 
\end{equation} where  $\chi_j(y)=(y^{2j+1}-y^{-2j-1})/(y-y^{-1})$ is the character
for the spin representation $j$ under $SU(2)$. Notice that 
we will absorb the minus sign
$(-1)^{2(j_L+j_R)}$ into the definition of $N^{\beta}_{j_L,j_R}$ below.

Combining all the contributions and taking into account the degeneracy $N^{\beta}_{j_L,j_R}$, we arrive at the 
following formula for the full nonperturbative refined topological string free energy $\mathcal{F}_{\text{ref,full}}$,  
\begin{equation}\label{eq:ref_full}
	\mathcal{F}_{\text{ref,full}}=\sum_{\stackrel{\beta, j_L, j_R}{n \in \IZ}} N^{\beta}_{j_L,j_R} \int_{\epsilon}^{\infty} \frac{ds}{s} \frac{e^{-s |z_{\beta,n}|^2}\ 
		\chi_{j_L}(e^{i \bar{z}_{\beta,n} s\lambda(\bb+\bb^{-1})/2})\ \chi_{j_R}(e^{i \bar{z}_{\beta,n} s\lambda(\bb-\bb^{-1})/2})}{4 \sin (\bar{z}_{\beta,n} s \bb \lambda /2)\sin (\bar{z}_{\beta,n} s \lambda / 2\bb)} \ ,
\end{equation} where $N^{\beta}_{j_L,j_R}$ count BPS states with class $\beta$ and spin $(j_L, j_R)$ under the little group $SU(2)_L \times SU(2)_R$.

\section{Comparison with \cite{AMP24}}

\subsection{Contour integral formula for refined topological string} Now we analyze the formula (\ref{eq:ref_full}). Denote the complexified K\"ahler form as
\begin{equation}
	\bt = {\bf v} + i {\bf B},
\end{equation} where ${\bf v}$ is ample. Recall the relation
\begin{equation}
	z_{\beta,n} = \beta \cdot \bt - 2 \pi i n = 
	\bd \cdot \bt - 2 \pi i n.
\end{equation}

Given a choice of $\bd, j_L, j_R$, perform a change of
variable 
\begin{equation}
	u_n = s \bar{z}_{\beta,n} = u.
\end{equation}

First let us express (\ref{eq:ref_full}) in the following form
\begin{equation}
	\mathcal{F}_{\text{ref,full}} = 
	\sum_{\beta, j_L, j_R} N^{\beta}_{j_L,j_R} \mathcal{F}_{\bd, j_L, j_R}, 
\end{equation}
where $\mathcal{F}_{\bd, j_L, j_R}$ is given by
\begin{equation}\label{eq:sum1}
\mathcal{F}_{\bd, j_L, j_R} = \sum_{n \in \IZ} \int_{0^+}^{\infty e^{i\theta_n}} \frac{du}{u} 
\frac{e^{-u z_n}
\chi_{j_L}(e^{i u \lambda(\bb+\bb^{-1})/2})\ 
\chi_{j_R}(e^{i u \lambda(\bb-\bb^{-1})/2})}
{4 \sin (u \bb \lambda /2)\sin (u \lambda / 2 \bb )},
\end{equation} where 
\begin{equation}
	z_n= \bd \cdot \bt - 2 \pi i n, \quad 
	\theta_n = \arctan\Big(\frac{2\pi n - \bd \cdot {\bf B} }{\bd \cdot {\bf v}}\Big).
\end{equation} 

First restrict the the case $\lambda \in \IR_+$. In this case
all the poles lie on the real axis, so that we can bring all 
integration contours with $2\pi n - \bd \cdot {\bf B}>0$ to lie right above the real axis and the one with $2\pi n - \bd \cdot {\bf B}<0$ to lie right below the real axis. 
Therefore we can write (\ref{eq:sum1}) as
\begin{align}\label{eq:sum2}
	\mathcal{F}_{\bd, j_L, j_R} &= \sum_{\stackrel{n \in \IZ}{2\pi n - \bd \cdot {\bf B}>0}} \int_{0^+}^{\infty e^{i 0^+}} \frac{du}{u} 
	\frac{e^{-u z_n}
		\chi_{j_L}(e^{i u \lambda(\bb+\bb^{-1})/2})\ 
		\chi_{j_R}(e^{i u \lambda(\bb-\bb^{-1})/2})}
	{4 \sin (u \bb \lambda /2)\sin (u \lambda / 2 \bb )}, \nonumber \\
	&+\sum_{\stackrel{n \in \IZ}{2\pi n - \bd \cdot {\bf B}<0}} \int_{0^+}^{\infty e^{i 0^-}} \frac{du}{u} 
	\frac{e^{-u z_n}
		\chi_{j_L}(e^{i u \lambda(\bb+\bb^{-1})/2})\ 
		\chi_{j_R}(e^{i u \lambda(\bb-\bb^{-1})/2})}
	{4 \sin (u \bb \lambda /2)\sin (u \lambda / 2 \bb )} .
\end{align}

Define $n_{\bd \cdot {\bf B}}$ by
\begin{equation}
	\bd \cdot {\bf B} = 2 \pi n_{\bd \cdot {\bf B}} + 
	\widetilde{\bd \cdot {\bf B}},
\end{equation} where $n_{\bd \cdot {\bf B}} = \lfloor \frac{\bd \cdot {\bf B} }{2\pi} \rfloor$ and $\widetilde{\bd \cdot {\bf B}} \in [0,2\pi)$.

We have $|e^{2\pi i u}|<1$ for $u$ deformed into upper half plane, and $|e^{-2\pi i u}|<1$ for $u$ deformed into lower half plane. Therefore we have 
\begin{align}
	&\sum_{\stackrel{n \in \IZ}{2\pi n - \bd \cdot {\bf B}>0}}
	e^{-u z_n}
	= \frac{e^{-(\bd \cdot \bt - 2 \pi i (n_{\bd \cdot {\bf B}}+1))u }}{1-e^{2\pi i u}} 
	= - \frac{e^{-(\bd \cdot \bt - 2 \pi i n_{\bd \cdot {\bf B}} )u }}{1-e^{-2\pi i u}}, \\
	&\sum_{\stackrel{n \in \IZ}{2\pi n - \bd \cdot {\bf B}<0}}
	e^{-u z_n} = \frac{e^{-(\bd \cdot \bt - 2 \pi i n_{\bd \cdot {\bf B}} )u }}{1-e^{-2\pi i u}}.
\end{align}	

Equation (\ref{eq:sum2}) now becomes
\begin{align}\label{eq:sum3}
	\mathcal{F}_{\bd, j_L, j_R} &= \Big(-\int_{0^+}^{\infty e^{i 0^+}} + \int_{0^+}^{\infty e^{i 0^-}} \Big)
	\frac{du}{u} \frac{e^{-(\bd \cdot \bt - 2 \pi i n_{\bd \cdot {\bf B}} )u }}{1-e^{-2\pi i u}}
	\frac{\chi_{j_L}(e^{i u \lambda(\bb+\bb^{-1})/2})\ 
		\chi_{j_R}(e^{i u \lambda(\bb-\bb^{-1})/2})}
	{4 \sin (u \bb \lambda /2)\sin (u \lambda / 2 \bb )}, \nonumber \\
	& = \oint_C \frac{du}{u} 
	\frac{e^{-(\bd \cdot \bt - 2 \pi i n_{\bd \cdot {\bf B}} )u }}{1-e^{-2\pi i u}}
	\frac{ \chi_{j_L}(e^{i u \lambda(\bb+\bb^{-1})/2})\ 
		\chi_{j_R}(e^{i u \lambda(\bb-\bb^{-1})/2})}
	{4 \sin (u \bb \lambda /2)\sin (u \lambda / 2 \bb )}, 
\end{align} where the contour $C$ circles all the poles 
on the positive real axis counterclockwisely.  

Summing over $\bd, j_L, j_R$, we have the full nonperturbative
refined topological string free energy,
\begin{equation}\label{eq:ref_full_contour}
	\mathcal{F}_{\text{ref,full}}=\sum_{\bd, j_L, j_R} N^{\bd}_{j_L,j_R} \oint_{C} \frac{du}{u} \frac{1}{1-e^{-2\pi i u}}
	\frac{e^{-(\bd \cdot \bt -2\pi i n_{\bd \cdot {\bf B}})u} \ 
		\chi_{j_L}(e^{i u \lambda(\bb+\bb^{-1})/2})\ \chi_{j_R}(e^{i u \lambda(\bb-\bb^{-1})/2})}{4 \sin (u \bb \lambda /2)\sin (u \lambda / 2 \bb )}, 
\end{equation} where $C$ is the contour $\Big(-\int_{0^+}^{\infty e^{i 0^+}} + \int_{0^+}^{\infty e^{i 0^-}} \Big)$. We propose that formula (\ref{eq:ref_full_contour}) is the full nonperturbative
refined topological string free energy, even if the coupling 
constant $\lambda$ is not real. 

When $\lambda$ is real, the contour integral over $C$ contains three families of poles,
\begin{equation}
	u \in \mathbb{N}_+, \quad u \in \frac{2\pi \bb}{\lambda} \mathbb{N}_+, \quad
	u \in \frac{2\pi}{\lambda\bb} \mathbb{N}_+,
\end{equation} where the first family gives the perturbative part in the refined free energy,

\begin{equation}\label{eq:ref_pert}
	\mathcal{F}_{\text{ref,pert}} = \sum_{k \in \mathbb{N}, {\bf d}} \sum_{j_L, j_R}	
	\frac{1}{k} \frac{ \chi_{j_L}(e^{i k \epsilon_L})\ \chi_{j_R}(e^{i k \epsilon_R})}{4 \sin ( k \lambda \bb /2) \sin ( k \lambda / 2\bb)} 
	N^{{\bf d}}_{j_L, j_R} e^{-k {\bf d} \cdot {\bf t}}.
\end{equation}
We note that (\ref{eq:ref_pert}) matches the refined Gopakumar-Vafa expansion presented in \cite{IKV09}, and is therefore also consistent with the index computation framework proposed in \cite{NO16}.

If we write $\mathcal{F}_{\text{ref,full}}$ as 
\begin{equation}
	\mathcal{F}_{\text{ref,full}} = \mathcal{F}_{\text{ref,pert}} + \mathcal{F}_{\text{ref,np}},
\end{equation} we see that the poles contributing to 
$\mathcal{F}_{\text{ref,np}}$, $\frac{2\pi \bb}{\lambda} l$ and 
$\frac{2\pi}{\lambda\bb}l$, reproduce the Borel singularities found in \cite{AMP24}.

\subsection{Stokes jumps}
To see the Stokes jump in (\ref{eq:ref_full_contour}), we need 
to bring the coupling constant $\lambda$ to be a complex number, as was investigated in \cite{HP24a}.
Let $\lambda = \lambda_{\text{re}} + i \lambda_{\text{im}}$, where we assume $\lambda_{\text{re}}>0$ for convenience.

Then the nonperturbative poles, $\frac{2\pi \bb}{\lambda} l$ and $\frac{2\pi}{\lambda\bb}l$ are rotated about the origin,
and lie on two half lines. Denote these two half lines by $\ell_{\bb \lambda}$ and $\ell_{\bb^{-1} \lambda}$. Also assume that $\bb$ 
is small so that these two lines are close enough. Denote
the angles of $\ell_{\bb \lambda}$ and $\ell_{\bb^{-1} \lambda}$
by $\theta_{\bb \lambda}$ and $\theta_{\bb^{-1} \lambda}$
respectively.

If $\theta_{n_{\bd \cdot {\bf B}}+1} > \theta_{\bb \lambda} > \theta_{n_{\bd \cdot {\bf B}}}$
and $\theta_{n_{\bd \cdot {\bf B}}+1} > \theta_{\bb^{-1} \lambda} > \theta_{n_{\bd \cdot {\bf B}}}$, 
then the result is still given by the expression (\ref{eq:ref_full_contour}) in which $C$ becomes 
the contour circling the poles on the positive real axis, and two half lines $\ell_{\bb \lambda}$ and $\ell_{\bb^{-1} \lambda}$.

Now assume
\begin{align}
	\theta_{n_{\bd \cdot {\bf B}+k+1}} > \theta_{\bb \lambda} > \theta_{n_{\bd \cdot {\bf B}+k}}, \nonumber \\ 
	\theta_{n_{\bd \cdot {\bf B}+k+1}} > \theta_{\bb^{-1} \lambda} > \theta_{n_{\bd \cdot {\bf B}+k}} .
\end{align}

Comparing the difference between the following two integrals, 
\begin{align}
	&\oint_{\widetilde{C}} \frac{du}{u} \frac{1}{1-e^{-2\pi i u}}
	e^{-(\bd \cdot \bt -2\pi (n_{\bd \cdot {\bf B}}+k))u} \cdots,
	\nonumber \\
	&\oint_{\widetilde{C}} \frac{du}{u} \frac{1}{1-e^{-2\pi i u}}
    e^{-(\bd \cdot \bt -2\pi n_{\bd \cdot {\bf B}})u} \cdots
\end{align} where $\widetilde{C}$ surrounds all the poles on
the positive real axis and two half lines 
$\ell_{\bb \lambda}$ and $\ell_{\bb^{-1} \lambda}$,
we obtain 
\begin{align}\label{eq:stokesjump1}
	\mathcal{F}_{\text{ref,full}} &= \sum_{\bd, j_L, j_R} N^{\bd}_{j_L,j_R} \oint_{\widetilde{C}} 
	\frac{du}{u} \frac{1}{1-e^{-2\pi i u}}
	\frac{e^{-(\bd \cdot \bt -2\pi i n_{\bd \cdot {\bf B}})u} \ 
	\chi_{j_L}(e^{i u \lambda(\bb+\bb^{-1})/2})\ \chi_{j_R}(e^{i u \lambda(\bb-\bb^{-1})/2})}{4 \sin (u \bb \lambda /2)\sin (u \lambda / 2 \bb )} \nonumber \\
	&+\sum_{w=1}^{k}
	\sum_{\bd, j_L, j_R} N^{\bd}_{j_L,j_R} \oint_{\widetilde{C}_{\lambda}} 
	\frac{du}{u}
	\frac{e^{-(\bd \cdot \bt -2\pi i n_{\bd \cdot {\bf B}} 
			-2 \pi i w)u} \ 
	\chi_{j_L}(e^{i u \lambda(\bb+\bb^{-1})/2})\ \chi_{j_R}(e^{i u \lambda(\bb-\bb^{-1})/2})}{4 \sin (u \bb \lambda /2)\sin (u \lambda / 2 \bb )} 
\end{align} where $\widetilde{C}_{\lambda}$ circles around 
the poles on $\ell_{\bb \lambda}$ and $\ell_{\bb^{-1} \lambda}$.

Fixing $w \in \{1,2,\cdots,k\}$,
we have 
\begin{equation}
	\bd \cdot \bt -2\pi i n_{\bd \cdot {\bf B}} 
	-2 \pi i w = \bd \cdot \bt + 2 \pi i n = \frac{A_{\bd,n}}{2\pi},
\end{equation} for some $n$.

The contribution of this particular value of $w$ to 
the contour integral along $\widetilde{C}_{\lambda}$ in (\ref*{eq:stokesjump1}) comes from the poles
\begin{equation}
	u = \frac{2\pi \bb}{\lambda} l,\quad \frac{2\pi}{\lambda\bb} l, \quad l\in \mathbb{N}_+.
\end{equation}
Collecting the contributions from these poles,
we have 
\begin{align}
	&\sum_{\bd, j_L, j_R} N^{\bd}_{j_L,j_R} \oint_{\widetilde{C}_{\lambda}} 
	\frac{du}{u}
	\frac{e^{-(\bd \cdot \bt -2\pi i n_{\bd \cdot {\bf B}} 
			-2 \pi i w)u} \ 
		\chi_{j_L}(e^{i u \lambda(\bb+\bb^{-1})/2})\ \chi_{j_R}(e^{i u \lambda(\bb-\bb^{-1})/2})}{4 \sin (u \bb \lambda /2)\sin (u \lambda / 2 \bb )} 
		\nonumber \\
	&= \sum_{\bd, j_L, j_R} N^{\bd}_{j_L,j_R}
	\sum_{l \geq 1} \frac{(-1)^l}{l}\Big[\frac{ \chi_{j_L}(\tilde{y}_\bb^l)\ \chi_{j_R}(\tilde{y}_\bb^l)}{2 \sin ( \pi l / \bb^2)} e^{-\frac{l A_{\bd,n}}{\bb \lambda}} + 
	\frac{ \chi_{j_L}(y_\bb^l)\ \chi_{j_R}(y_\bb^l)}{2 \sin ( \pi l \bb^2)} e^{-\frac{l \bb A_{\bd,n}}{\lambda}} \Big] 
	\nonumber \\
	&= i \sum_{\bd, j_L,j_R} N^{{\bf d}}_{j_L, j_R} \log \Phi_\bb^{[j_L,j_R]}
	\Big( -\frac{A_{\bd,n}}{2 \pi \lambda}\Big) = i \sum_{\bd, j} 
	\Omega_{[j]}(\bd) \log \Phi_\bb^{[j]} \Big( -\frac{A_{\bd,n}}{2 \pi \lambda}\Big),
\end{align} which is one of the summands in (\ref{eq:tseries2}) and coincides with the trans-series solutions
in \cite{AMP24}, up to a minus sign.

\section{Conclusion and discussion}

In this paper we propose a new formula (\ref{eq:ref_full_contour}) for the full nonperturbative refined topological string free energy, extending the results in \cite{HP24a}\cite{HP24b}. 

Some remarks on this formula are in order. The perturbative
part $\mathcal{F}_{\text{ref,pert}}$ of the free energy 
(\ref{eq:ref_pert}) is consistent with the results in \cite{IKV09}\cite{CDP14}.

Summing over the nonperturbative poles in the formula,
we reproduce the trans-series and Stokes automorphisms discussed in \cite{AMP24}.

This formulation also raises several intriguing questions. 
In our derivation of the Stokes jumps, we obtain the 
Stokes automorphisms as certain wallcrossings with respect 
to the string coupling $\lambda$ deformed into a complex number. Another way to see the Stoke automorphisms is to 
use the so-called dual partition function,
which is a discrete Fourier transform of the usual 
partition function. The dual partition function also appears as the isomonodromic tau function \cite{CPT23}. 
Additionally, it is conjectured that the full nonperturbative
free energy is related to the tau function, restricted along 
a specific slice in the space of $(\bt, \widetilde{\bt})$, where 
 $\widetilde{\bt}$ is the dual coordinates of $\bt$. Therefore this 
 raises the question of what new insights into the tau function 
 might be obtained from the formula in this paper or the results in \cite{HP24a}. Finally the works by Bridgeland \cite{Bri1}
 \cite{Bri2} suggest the existence of the tau function, as
 the generating function for the solution to a class of Riemann-Hilbert problems. It would be interesting to explore whether the full nonperturbative free energy provides any novel perspectives on these solutions.

\medskip

\noindent{\bf Acknowledgements} We would like to thank the referee whose comments helped improve
greatly the presentation of the paper. WYC was partially supported by Taiwan NSTC grant 112-2115-M-002-007-MY2 and NTU Core Consortiums grant 113L893603 113L893604 (TIMS). 

\medskip

\noindent{\bf Data availability} No datasets were generated or analysed during the current study.

\medskip\medskip

\noindent{\Large \bf Declaration}

\medskip

\noindent{\bf Conflict of interest} The author declares no conflict of interest.
 
\bibliographystyle{alpha}

\end{document}